\documentclass{article}
\usepackage{spconf,amsmath,graphicx,xcolor,booktabs,hyperref, pifont}
\usepackage{tikz}
\usepackage{enumitem}
\usepackage{multirow}
\usepackage{subcaption}
\usetikzlibrary{
  arrows.meta,
  positioning,
  calc,
  fit,
  backgrounds
}

\def\x{{\mathbf x}}

\newcommand{\cmark}{\ding{51}}
\newcommand{\xmark}{\ding{55}}
\newcommand{\field}[1]{\texttt{#1}}
\newcommand{\qtype}[1]{\textsc{#1}}
\newcommand{\figcheck}{\textcolor{green!55!black}{\cmark}}
\newcommand{\figcross}{\textcolor{red!75!black}{\xmark}}
\newcommand{\musicwave}{\tikz[baseline=0pt]{\foreach \x/\h in {0/2,3/4,6/3,9/5,12/2.5,15/4.5,18/2,21/3.5,24/5,27/3,30/4.5,33/2.5,36/4,39/2,42/3.5,45/2.5}{\draw[teal!55!black,line width=1.1pt] (\x pt,-\h pt) -- (\x pt,\h pt);}}}
\title{Text-to-music Generation via Structured Captions AND Interpretable Evaluation}

\name{\shortstack{Junkai Wu$^{\sharp\flat*}$ \quad Zhepei Wang$^{\flat}$ \quad Ge Zhu$^{\flat}$ \quad Sarah Hye-yeon Lee$^{\flat}$ \\[-2pt]
Mari Ostendorf$^{\sharp}$ \quad Nicholas J. Bryan$^{\flat}$}
\thanks{$^{*}$~Work done while an intern at Adobe Research.} }
\address{$^{\sharp}$\,University of Washington \qquad $^{\flat}$\,Adobe Research}

\begin{document}
\ninept
\maketitle
\begin{abstract}
Text descriptions of music are central to the performance of text-to-music generation (TTM) methods, but creation of paired text information for music generation is relatively underexplored. Here, we propose training latent diffusion/flow-based music generation models conditioned on structured captions (SCAPs) and compare with typical free-form captions via a set of controlled ablations.
We compare free-form vs.\ structured captions, partial vs.\ full information coverage, and global vs.\ temporal text information during training, by fine-tuning different configurations of a pretrained flow-based diffusion transformer (DiT) and using a multi-stage inference-time prompt re-writer. Further, we introduce a new text-alignment evaluation protocol
inspired by Davidsonian Scene Graphs (DSGs) for a fine-grained, interpretable measure of faithfulness of TTM methods. With multiple metrics, we show 1) that fine-tuning our DiT on SCAPs significantly improves text-alignment on music generation outputs while also improving quality, and 2) our proposed Music DSG metric better correlates with human preference than CLAP.
\end{abstract}

\section{Introduction}
\label{sec:intro}


Text-to-music (TTM) generation aims to synthesize music faithful to the input prompt and demonstrates strong acoustic and aesthetics quality. Conventional TTM methods are trained with a large corpus of music paired with textual descriptions that include musical attributes such as genre, instrumentation, mood, tempo, keywords, and more. Recent TTM models include autoregressive methods based on language models over audio tokens such as MusicLM~\cite{musiclm}, MusicGen~\cite{musicgen}, and YuE~\cite{yue}, methods based on diffusion or flow-matching, such as Stable Audio Open~\cite{sao, sao3} and MusicFlow~\cite{musicflow}, and hybrid methods that combine language model planning and diffusion rendering, such as Seed-Music~\cite{seed-music}, ACE-Step~\cite{acestep15}, and Qwen-Music~\cite{qwenmusic}. Most of the prior work has focused on improved model architecture, audio representation, training scale, and sampling algorithms.

In contrast, synthetic data curation and the structure of paired textual information used for conditioning has received less investigation. Questions such as how much musical information is contained in training captions (full vs. partial information), how that information is formatted (free-form vs. explicit structure), how text data is managed over time via global or time-varying textual information (strong- vs.\ weakly-labeled captioning), if inference-time prompt rewriting, enhancement, and/or formatting is needed, and how best to evaluate text alignment. These choices affect both acoustic quality and prompt adherence. In text-to-image generation, DALL$\cdot$E~3 shows improved image generation quality by training with detailed synthetic captions~\cite{dalle3}, while PixArt-$\alpha$ emphasizes the benefits of training with dense, high-information captions \cite{pixart-alpha}. Additionally, in text-to-video generation, structured captions (SCAPs) have also been explored to represent entities, attributes and motion more explicitly. The image and video results motivate treating the text side of paired TTM training data as an important modeling choice. 

\begin{figure*}[!t]
  \centering
  \resizebox{\textwidth}{!}{%
  \begin{tikzpicture}[
    x=1cm,
    y=1cm,
    font=\sffamily\footnotesize,
    >={Latex[length=1.8mm,width=1.6mm]},
    flow/.style={
      -{Latex[length=1.8mm,width=1.6mm]},
      line width=0.68pt,
      draw=black!62
    },
    stage/.style={
      rounded corners=3pt,
      draw=black!52,
      line width=0.52pt,
      align=center,
      inner xsep=5pt,
      inner ysep=4pt,
      fill=white
    },
    gcard/.style={
      stage,
      fill=orange!10,
      draw=orange!65!black,
      text width=1.75cm
    },
    tcard/.style={
      stage,
      fill=cyan!9,
      draw=cyan!60!black,
      text width=2.00cm
    },
    flab/.style={
      font=\tiny\itshape,
      text=black!62,
      fill=white,
      inner sep=0.8pt,
      align=center
    },
    ptitle/.style={
      font=\sffamily\bfseries\small,
      anchor=west
    }
  ]

  \begin{scope}[on background layer]
    \fill[black!1.2, rounded corners=4pt]
      (0.00,0.00) rectangle (7.45,8.05);
    \draw[black!18, rounded corners=4pt, line width=0.45pt]
      (0.00,0.00) rectangle (7.45,8.05);

    \fill[black!1.2, rounded corners=4pt]
      (7.75,0.00) rectangle (12.95,8.05);
    \draw[black!18, rounded corners=4pt, line width=0.45pt]
      (7.75,0.00) rectangle (12.95,8.05);

    \fill[black!1.2, rounded corners=4pt]
      (13.25,0.00) rectangle (19.65,8.05);
    \draw[black!18, rounded corners=4pt, line width=0.45pt]
      (13.25,0.00) rectangle (19.65,8.05);
  \end{scope}

  \node[ptitle] at (0.26,7.72) {(a) Training};

  \node[
    stage,
    fill=blue!7,
    minimum width=6.78cm,
    minimum height=0.62cm
  ] (db) at (3.72,7.00)
    {\textbf{Music track \& metadata}};

  \node[
    stage,
    fill=blue!4,
    minimum width=1.22cm,
    minimum height=0.66cm,
    inner xsep=2pt,
    inner ysep=2pt
  ] (crop1) at (1.02,5.78)
    {\scriptsize\textbf{Crop 1}\\[-2pt]
     \scalebox{0.58}{\musicwave}};

  \node[
    stage,
    fill=blue!4,
    minimum width=1.22cm,
    minimum height=0.66cm,
    inner xsep=2pt,
    inner ysep=2pt
  ] (crop2) at (1.02,4.78)
    {\scriptsize\textbf{Crop 2}\\[-2pt]
     \scalebox{0.58}{\musicwave}};

  \node[
    stage,
    fill=blue!4,
    minimum width=1.22cm,
    minimum height=0.66cm,
    inner xsep=2pt,
    inner ysep=2pt
  ] (crop3) at (1.02,3.78)
    {\scriptsize\textbf{Crop 3}\\[-2pt]
     \scalebox{0.58}{\musicwave}};

  \node[tcard, minimum height=0.58cm] (t1) at (3.72,5.78)
    {$\mathbf{TSCAP}_{\scriptscriptstyle 0\text{--}30\,\mathrm{s}}$\\[-3pt]
     \scriptsize \field{genre}: \textcolor{blue!70!black}{\ldots}\quad
     \field{mood}: \textcolor{blue!70!black}{\ldots}\quad \ldots};

  \node[tcard, minimum height=0.58cm] (t2) at (3.72,4.78)
    {$\mathbf{TSCAP}_{\scriptscriptstyle 10\text{--}40\,\mathrm{s}}$\\[-3pt]
     \scriptsize \field{genre}: \textcolor{blue!70!black}{\ldots}\quad
     \field{mood}: \textcolor{blue!70!black}{\ldots}\quad \ldots};

  \node[tcard, minimum height=0.58cm] (t3) at (3.72,3.78)
    {$\mathbf{TSCAP}_{\scriptscriptstyle 20\text{--}50\,\mathrm{s}}$\\[-3pt]
     \scriptsize \field{genre}: \textcolor{blue!70!black}{\ldots}\quad
     \field{mood}: \textcolor{blue!70!black}{\ldots}\quad \ldots};

  \node[font=\sffamily\bfseries\normalsize, text=black!58]
    (tdots) at (3.72,3.17) {\ldots};

  \begin{scope}[on background layer]
    \node[
      draw=cyan!42!black,
      fill=cyan!2,
      rounded corners=3pt,
      line width=0.38pt,
      inner xsep=3.5pt,
      inner ysep=2.6pt,
      fit=(t1)(t2)(t3)(tdots)
    ] (tgroup) {};
  \end{scope}

  \node[gcard, minimum height=0.58cm] (gscap) at (6.18,4.78)
    {\textbf{GSCAP}\\[-3pt]
     \scriptsize \field{genre}: \textcolor{blue!70!black}{\ldots}\quad
     \field{mood}: \textcolor{blue!70!black}{\ldots}\quad \ldots};

  \node[
    stage,
    fill=teal!7,
    text width=3.08cm,
    minimum height=0.62cm
  ] (conditioning) at (3.72,1.86)
    {\textbf{SCAP selection}\\[-2pt]
     \scriptsize source: G/T \quad coverage: F/P};

  \node[
    stage,
    fill=violet!9,
    draw=violet!58!black,
    minimum width=3.36cm,
    minimum height=0.70cm
  ] (trainDit) at (3.72,0.58)
    {\textbf{SCAP-conditioned DiT}};

  \coordinate (cropStemTop) at (0.22,6.42);
  \coordinate (cropStemBottom) at (0.22,3.78);
  \draw[line width=0.68pt, draw=black!62]
    ($(db.south west)+(0.18,0)$) -- (cropStemTop) -- (cropStemBottom);
  \draw[flow] (0.22,5.78) -- (crop1.west);
  \draw[flow] (0.22,4.78) -- (crop2.west);
  \draw[flow] (0.22,3.78) -- (crop3.west);
  \draw[flow] (db.south -| tgroup.north) -- (tgroup.north);
  \draw[flow] (db.south -| gscap.north) -- (gscap.north);

  \draw[flow] (crop1.east) -- (t1.west);
  \draw[flow] (crop2.east) -- (t2.west);
  \draw[flow] (crop3.east) -- (t3.west);

  \draw[flow] (tgroup.south) -- (conditioning.north);

  \coordinate (gturn) at (gscap.south |- conditioning.east);
  \draw[flow]
    (gscap.south) -- (gturn) -- (conditioning.east);

  \draw[flow] (conditioning) -- (trainDit);

  \node[ptitle] at (8.02,7.72) {(b) Inference};

  \node[
    stage,
    fill=gray!5,
    text width=4.35cm,
    minimum height=0.74cm
  ] (prompt) at (10.35,6.66)
    {\textbf{Free-form prompt}\\[-1pt]
     \itshape ``upbeat lo-fi hip-hop with piano''};

  \node[
    stage,
    fill=blue!8,
    draw=blue!55!black,
    text width=4.35cm,
    minimum height=1.02cm
  ] (transcribe) at (10.35,5.05)
    {\textcolor{blue!72!black}{\textbf{Transcribe: partial SCAP}}\\[-1pt]
     \scriptsize
     \field{genre}: \textcolor{blue!72!black}{lo-fi hip-hop}\quad
     \field{mood}: \textcolor{blue!72!black}{upbeat}\\[-1pt]
     \field{inst}: \textcolor{blue!72!black}{piano}\quad
     \field{bpm}: \phantom{82}\quad
     \field{key}: \phantom{F\,min}};

  \node[
    stage,
    fill=red!7,
    draw=red!55!black,
    text width=4.35cm,
    minimum height=1.02cm
  ] (infer) at (10.35,3.50)
    {\textcolor{red!72!black}{\textbf{Infer: completed SCAP}}\\[-1pt]
     \scriptsize
     \field{genre}: \textcolor{blue!72!black}{lo-fi hip-hop}\quad
     \field{mood}: \textcolor{blue!72!black}{upbeat}\\[-1pt]
     \field{inst}: \textcolor{blue!72!black}{piano}\quad
     \field{bpm}: \textcolor{red!72!black}{82}\quad
     \field{key}: \textcolor{red!72!black}{F\,min}};

  \node[
    stage,
    fill=violet!9,
    draw=violet!58!black,
    minimum width=3.62cm,
    minimum height=0.70cm
  ] (inferDit) at (10.35,2.12)
    {\textbf{SCAP-conditioned DiT}};

  \node[
    stage,
    fill=orange!13,
    minimum width=3.62cm,
    minimum height=0.74cm
  ] (music) at (10.35,0.80)
    {\textbf{Generated music}\\[2pt]\musicwave};

  \draw[flow] (prompt) -- (transcribe);
  \draw[flow] (transcribe) -- (infer);
  \draw[flow] (infer) -- (inferDit);
  \draw[flow] (inferDit) -- (music);

  \node[ptitle] at (13.52,7.78) {(c) Music DSG evaluation};

  \node[
    stage,
    fill=gray!5,
    text width=2.55cm,
    minimum height=0.78cm
  ] (evalPrompt) at (14.85,6.66)
    {\textbf{Free-form prompt}\\[-1pt]
     \itshape ``F-key fast-tempo lo-fi music with piano''};

  \node[
    stage,
    fill=orange!13,
    text width=2.40cm,
    minimum height=0.78cm
  ] (evalMusic) at (18.05,6.66)
    {\textbf{Generated music}\\[2pt]\musicwave};

  \node[
    stage,
    fill=blue!6,
    draw=blue!48!black,
    text width=2.82cm,
    minimum height=1.16cm
  ] (decompose) at (16.45,4.25)
    {\scriptsize\textbf{Text decomposition}\\[-1pt]
     \scriptsize lo-fi? \quad key F?\\[-1pt]
     piano? \quad fast?\\[-1pt]
     };

  \node[
    stage,
    fill=violet!7,
    draw=violet!48!black,
    text width=2.82cm,
    minimum height=1.16cm
  ] (qa) at (16.45,2.60)
    {\textbf{Music Q\&A}\\[-1pt]
     \scriptsize lo-fi? \figcheck\quad key F? \figcross\\[-1pt]
     piano? \figcheck\quad fast? \figcheck\\[-1pt]
     };

  \node[
    font=\sffamily\bfseries\scriptsize,
    text=violet!65!black
  ] (dsgtitle) at (16.45,5.30) {Music DSG};

  \begin{scope}[on background layer]
    \node[
      draw=violet!42!black,
      fill=violet!1.5,
      rounded corners=4pt,
      line width=0.55pt,
      inner xsep=6pt,
      inner ysep=7pt,
      fit=(dsgtitle)(decompose)(qa)
    ] (dsgbox) {};
  \end{scope}

  \node[
    stage,
    fill=green!7,
    draw=green!45!black,
    minimum width=2.65cm,
    minimum height=0.76cm
  ] (score) at (16.45,0.78)
    {\textbf{Faithfulness score}\quad 0.75};

  \coordinate (promptStart) at ($(evalPrompt.south)+(-0.50,0)$);
  \coordinate (promptTurn) at (promptStart |- decompose.west);
  \draw[flow] (promptStart) -- (promptTurn) -- (decompose.west);

  \coordinate (musicStart) at ($(evalMusic.south)+(0.50,0)$);
  \coordinate (musicTurn) at (musicStart |- qa.east);
  \draw[flow] (musicStart) -- (musicTurn) -- (qa.east);

  \draw[flow] (decompose.south) -- (qa.north);
  \draw[flow] (dsgbox.south) -- (score.north);

  \end{tikzpicture}%
  }

  \begin{subcaptiongroup}
    \phantomsubcaption\label{fig:pipeline-training}
    \phantomsubcaption\label{fig:pipeline-inference}
    \phantomsubcaption\label{fig:pipeline-evaluation}
  \end{subcaptiongroup}

  \caption{Overview of training, inference, and evaluation. During training (a), sampled crops are paired with their corresponding TSCAPs, while a
  track-level GSCAP provides global conditioning for DiT fine-tuning. At inference (b), Transcribe extracts
  explicit prompt information (shown in blue) and Infer completes the missing fields (in red) before music generation. For Music DSG evaluation (c), the prompt is decomposed
  into atomic questions, the generated music is verified by music Q\&A, and
  the answers are aggregated into a faithfulness score.}
  \label{fig:pipeline}
\end{figure*}
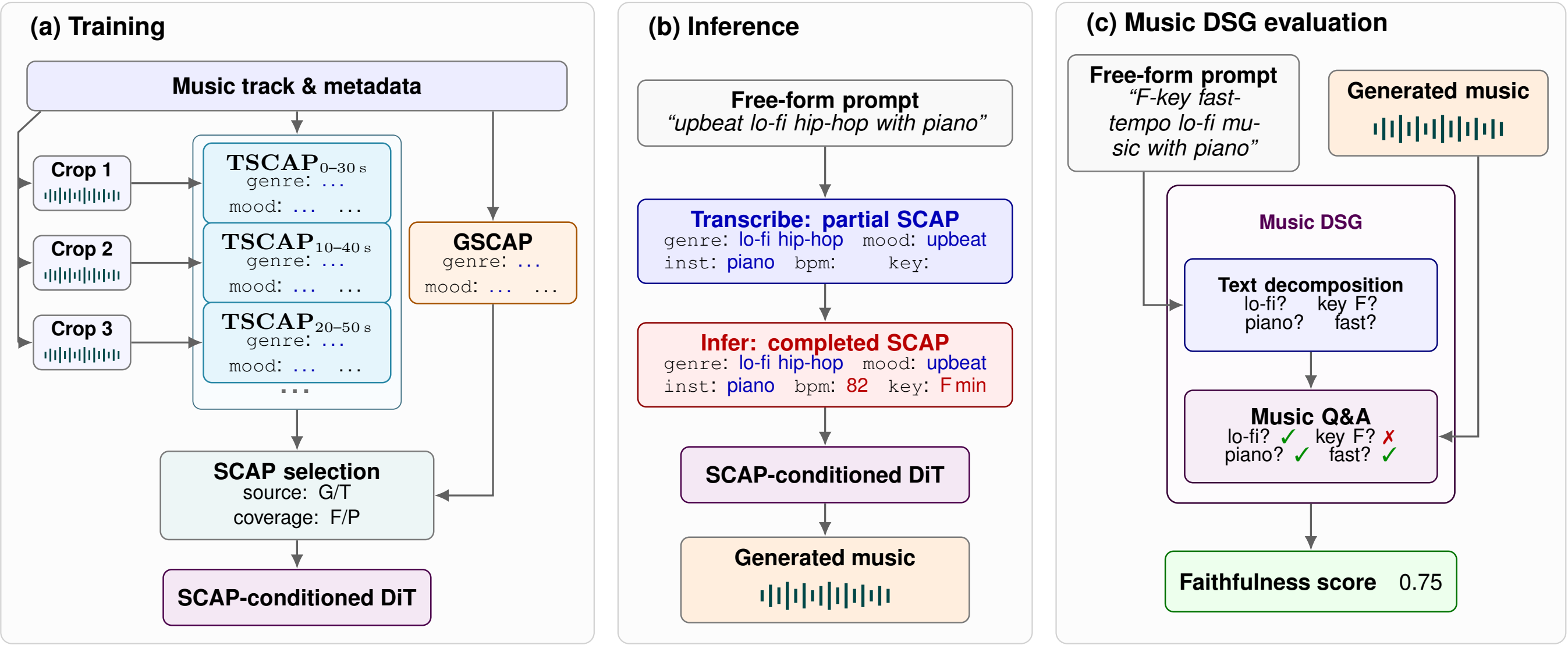

We propose \emph{structured captions} as conditioning for a latent diffusion/flow-based music generation~\cite{rectified-flow, scale-rectified-flow} 
, shown in Fig.~\ref{fig:pipeline}\subref{fig:pipeline-training}.
With this framework, we have three key contributions.
First, we compare free-form and structured conditioning and conduct a controlled study of attribute coverage (partial vs.\ full) and temporal granularity (global vs.\ time-varying) as independent caption design choices. 
Second, we introduce a multi-stage inference-time prompt re-writer that formats free-form user prompts into incomplete SCAPs and then enhances them to complete SCAPs fields while preserving user's intent, shown in Fig.~\ref{fig:pipeline}\subref{fig:pipeline-inference}. These ablations allow us to isolate how caption structure, richness of musical details, and temporal granularity affect generation quality and prompt adherence. 
Lastly, we introduce a new text-alignment evaluation protocol in Fig.~\ref{fig:pipeline}\subref{fig:pipeline-evaluation}, deemed Music Davidsonian Scene Graph (DSG)~\cite{dsg}, as a fine-grained and interpretable measure of faithfulness of TTM systems, and we empirically show that it correlates more strongly with listener preference than CLAP. We find SCAP-conditioned training across caption sources and coverage settings improves both text-alignment and audio quality for music generation. 
\footnote{Project page \url{https://mu-scap.github.io/}}
\vspace{-2mm}
\section{Related Work}


\vspace{-2mm}
\subsection{Text conditioning and prompt rewriting}
\label{ssec:related-text}

Prior work has explored improving textual supervision for audio and music generation and understanding. 
MusTango~\cite{mustango} augments music descriptions with chords, beats, and key; FUTGA~\cite{futga} synthesizes temporally structured, segment-level descriptions for
full-length music; Music Metadata LLMs~\cite{music-metadata-llm} represent music as structured metadata, convert it to free-form captions, and impute missing fields. These studies demonstrate the value of informative and temporally grounded music descriptions, but do not isolate the effects of caption structure, attribute coverage, and temporal granularity.

Recent methods also optimize prompts at inference. ACE-Step 1.5~\cite{acestep15} expands short queries into detailed song blueprints with metadata, lyrics, and long-form captions, while
Qwen-Music~\cite{qwenmusic} rewrites free-form prompts into structured musical tags and lyrics before semantic generation and acoustic rendering.
Our proposed rewriter differs by 
separating the faithful extraction of user-specified information into a partial SCAP from the completion of unspecified fields with musically compatible details. This design allows us to preserve the user's stated intent, while matching the structured caption distribution seen by the music generator during training.

\vspace{-3.9mm}
\subsection{Automatic music--text alignment evaluation}
\label{sec:related-eval}

Music-text alignment is commonly evaluated by similarity in a joint audio-text embedding space, such as CLAP~\cite{laion-clap}. MusicEval~\cite{musiceval} collects expert ratings of generated music to train a CLAP-based predictor of musicality and text alignment. CMI-RM~\cite{cmi} uses a transformer reward model to assess musicality, text alignment, and compositional instructions. These evaluators provide a holistic score but not detailed, interpretable attribute-level verification results. Rubric-guided LLM evaluation has been explored for conversational music recommendation, where an LLM judges system responses using user context and track metadata~\cite{doh2026llmjudge}; in contrast, our setting evaluates generated audio against the input prompt.

Question-answering metrics provide a complementary approach. TIFA~\cite{tifa} decomposes image prompts into question-answer pairs, while DSG~\cite{dsg}
adds semantic dependencies among atomic questions. For audio, AQAScore~\cite{aqascore} defines alignment as a probability on a set of audio question-answer pairs. AnyAudio-Judge~\cite{anyaudio} decomposes complex instructions into independently verifiable binary rubrics and evaluates speech, sound, and music generation. Similarly, AudioScope-TTA~\cite{audioscope-tta} derives fine-grained QA rubrics from structured soundscape annotations. Our proposed Music DSG specializes this approach to music attributes using question-specific verifiers based on neural networks or large audio language models
We explore both binary and confidence-based aggregation against human preferences.





\vspace{-3mm}
\section{Method}

\subsection{SCAP construction}
\label{sec:scap-construction}

Inspired by rich text representations for images~\cite{dalle3} and
music~\cite{music-metadata-llm}, we condition TTM models on structured
music captions (SCAPs).
A music SCAP is a natural way to build an information-rich caption with important musical attributes.
Each SCAP consists of a fixed set of fields: \field{genre}, \field{mood}, \field{energy}, \field{key}, \field{bpm}, \field{time signature}, \field{instruments}, \field{vocal}, and \field{keywords}. Information such as playing techniques, effects, temporal order, and dynamics that does not fit the other fields is preserved in \field{keywords}.
We construct SCAPs at two temporal granularities. A \emph{global SCAP (GSCAP)} contains track-level labels that characterize the full song and is shared by all training segments sampled from the track. A \emph{time-varying SCAP (TSCAP)} instead describes a local musical segment, representing temporal variation as a sequence of SCAPs. To obtain TSCAPs at scale, we fine-tune Qwen2.5-Omni~\cite{qwen25omni} on in-house track-level SCAP targets combining human-annotated metadata and MIR-inferred labels, and apply it to overlapping windows to produce TSCAP pseudo-labels.


\vspace{-4.5mm}
\subsection{SCAP-conditioned generator finetuning}
\label{sec:scap-dit}
We adapt a pretrained flow-based DiT from free-form text conditioning to SCAP conditioning as shown in Fig.~\ref{fig:pipeline}\subref{fig:pipeline-training}. We retain the generator architecture, latent audio representation, and flow-matching objective\cite{rectified-flow,scale-rectified-flow}. The primary change is that the free-form caption is replaced by GSCAP or the matched TSCAP constructed in Sec.~\ref{sec:scap-construction}. Hence, differences between the base and finetuned models can be attributed primarily to the textual conditioning.



We vary SCAP conditioning along two axes. The caption source is either
\emph{Global}, using a track-level GSCAP, or \emph{Temporal}, using a
local TSCAP. Independently, a \emph{Full} SCAP retains all available
fields, whereas a \emph{Partial} SCAP randomly drops a subset of fields
during training. Crossing source and coverage yields four configurations:
Full+Temporal (FT), Partial+Temporal (PT), Full+Global (FG), and
Partial+Global (PG). Their caption construction, audio-crop pairing, and
training setup are specified in Sec.~\ref{sec:dit_train}.

\subsection{Prompt-to-SCAP rewriter}
\label{sec:rewriter}

At inference, a two-stage prompt-to-SCAP rewriter converts a free-form user prompt into a complete SCAP suitable for conditioning the DiT. In Fig.~\ref{fig:pipeline}\subref{fig:pipeline-inference}, (1) \textbf{Transcribe} extracts only the musical information explicitly stated in the prompt and places it into the corresponding SCAP fields, producing a partial SCAP. Then, (2) \textbf{Infer} completes the SCAP by filling the remaining unspecified fields with musically compatible information that is consistent with the user's intent. This separation ensures that user-specified constraints are faithfully preserved before any missing information is inferred, reducing the risk that inferred values overwrite explicit intent. Both stages use training-free prompting of a state-of-the-art language model with broad musical knowledge. The rewriter ultimately produces the single SCAP that best represents the input prompt and matches the structured conditioning format expected by the generator. Unlike the training captions, its output has no TSCAP or GSCAP distinction.


\vspace{-3mm}
\subsection{Music DSG}
\label{sec:musicdsg}

Music DSG, as shown in Fig.~\ref{fig:pipeline}\subref{fig:pipeline-evaluation}, has three steps: (1) \textbf{Decomposition}: a decomposer turns a music caption into atomic yes/no questions of eleven types on two levels similar to~\cite{dsg}. The first level covers instrument/vocal/section \qtype{presence}, \qtype{genre}, \qtype{mood}, \qtype{energy}, \qtype{bpm}, \qtype{key}, \qtype{time signature}, and \qtype{exclusion/negation}. The second level covers \qtype{attribute}, \qtype{count}, and \qtype{relation}. Most second-level questions depend on one first-level parent, while a relation question can depend on multiple parents; for example, ``Does the piano come before the violin?'' depends on the presence questions for both instruments. A prompt can yield multiple questions of the same type, such as one presence question for each instrument mentioned. (2) \textbf{Music QA}: for each question, a music QA system returns 
a binary score (1 for ``yes'' and 0 for ``no'') and a confidence score (probability of ``yes'').
(3) \textbf{Scoring}: binary and confidence scores are averaged separately within each question type, then combined using a weighted average over the types present in the caption.
Before averaging, a second-level question's answer is set to ``no'' if any of its parent questions has answer ``no'', and its confidence is multiplied by the mean confidence of its parents.

\vspace{-2mm}
\section{Experiments}
\label{sec:experiments}

\subsection{SCAP-to-music training}
\label{sec:dit_train}

The base DiT (Base) is trained on randomly sampled 32-second music segments with free-form captions. We construct GSCAPs from track-level metadata in our internal music corpus. For GSCAP training (FG, PG), the track-level SCAP is paired with every segment sampled from the track. TSCAPs are obtained for every 30-second window with a 10-second hop. For TSCAP conditioning (FT, PT), each 32-second training crop is paired with the TSCAP window with maximum temporal overlap;  At long-form inference, each generated chunk is conditioned on the preceding audio, similar to~\cite{voicebox}.

We instantiate the FT, PT, FG, and PG configurations from Sec.~\ref{sec:scap-dit} on our internal music corpus, all initialized from Base as the free-form conditioning baseline. Each SCAP-conditioned model is fine-tuned for 150k steps on 32 A100 GPUs. The DiT architecture, latent audio representation, flow-matching objective, audio data, crop sampling, and previous chunk conditioning are held fixed. Thus, the controlled comparison varies only caption structure, field coverage, and temporal granularity.


\subsection{Prompt-to-SCAP development}
\label{sec:rewriter_dev}
For \textbf{Transcribe}, unless otherwise specified, we use Qwen3.5-35B to convert free-form prompts into partial SCAPs, filling fields other than \field{keywords} only when explicitly stated and placing remaining information in \field{keywords} as short phrases. For \textbf{Infer}, we compare zero-shot closed-source LLMs (Claude 4.8 Opus and GPT-5.6 Sol), open-source LLMs (Qwen3.5-35B, 112B, and 397B), and multimodal LLMs in text-only mode (FireRed Audio and Qwen3-Omni-30B). Claude and GPT perform both stages in their respective settings. We also include \textbf{Random}, which samples missing fields from the training data, and \textbf{Null}, which leaves the SCAP partial. Text LLMs may provide broad musical knowledge, while multimodal pretraining may provide broader exposure to music. We exclude Music Flamingo~\cite{ghosh2025music} because it does not reliably produce valid SCAPs.


\vspace{-3.5mm}
\subsection{Music DSG setup}
\label{sec:mdsg_setup}


We use Qwen3.5-35B-A3B for text decomposition and Qwen3-Omni-30B-A3B for Music QA, with in-house detectors for \qtype{key}, \qtype{bpm}, and \qtype{time signature}. The key detector follows S-KEY~\cite{skey}, and the BPM and time-signature detector uses a temporal convolutional beat tracker~\cite{davies2019tcn} with a dynamic Bayesian network post-filter~\cite{bock2016joint}. For LLM questions, confidence is defined as $p(\text{yes})/[p(\text{yes})+p(\text{no})]$;
detector questions use a graded score based on the distance from the requested value. Type weights are 1.5 for \qtype{presence}, \qtype{genre}, \qtype{mood}, \qtype{bpm}, and \qtype{exclusion/negation}; 1.0 for \qtype{energy}, \qtype{attribute}, and \qtype{count}; and 0.5 for \qtype{key}, \qtype{time signature}, and \qtype{relation}.

\vspace{-4mm}
\subsection{Evaluation}
\label{sec:eval}

\textbf{Music DSG validation.}
We measure the agreement between Music DSG and human preference using Spearman's $\rho$ on MusicEval benchmark~\cite{musiceval} and an internal dataset. MusicEval contains 413 generated clips from 129 prompts with expert musicality and alignment ratings; we use the utterance-level alignment ratings. Our internal set contains 600 clips from 200 prompts rendered by three TTM models. We collect pairwise alignment preferences from 128 general listeners over all 600 pairs (about 5 votes per pair) and 13 professional musicians over 400 pairs (about 3 votes per pair).

\smallskip
\noindent
\textbf{Music generation evaluation.} We evaluate the full music-generation pipeline on the Song Describer Dataset (SD)~\cite{song-describer}, using its instrumental-only subset of 586 clips with 4 generations per prompt, and genre9k, our internal set of 18 genres with 500 clips each. We report CLAP~\cite{laion-clap} and Music DSG for alignment, and FAD~\cite{fad} with a neural autoencoder backbone ~\cite{stae} and MuQ~\cite{muq-mulan} for audio quality. We also report the average relative improvement over Base (Avg.~$\Delta$) for CLAP, DSG and MuQ; for FAD, we report relative increment on the inverse for improvement.
\begin{table}[t]
  \centering
  \small
  \caption{Spearman correlation ($\rho$, higher is better) of automatic
  music--text alignment scores with human preference for Internal and MusicEval-test. 
DSG-conf and DSG-binary are our confidence-based and hard-$0/1$
variants; AnyAudio-Judge uses the 30B model.}
  \label{tab:dsg-correlation}
  \fontsize{9.}{10}\selectfont
  \setlength{\tabcolsep}{1.2pt}
  \renewcommand{\arraystretch}{0.84}
  \begin{tabular}{@{}l cc c@{}}
    \toprule
    Metric & Internal (general) & Internal (pro) & MusicEval \\
    \midrule
    \multicolumn{4}{@{}l}{\emph{Rubric verifiers}} \\
    \quad \textbf{DSG-conf} (ours)      & \textbf{0.421} & \textbf{0.414} & \textbf{0.248} \\
    \quad DSG-binary (ours)             & 0.351          & 0.334          & 0.193 \\
    \quad AnyAudio-Judge                & 0.414          & 0.385          & 0.225 \\
    \midrule
    \multicolumn{4}{@{}l}{\emph{Contrastive embeddings}} \\
    \quad CLAP      & 0.165 & 0.211 & 0.235 \\
    \quad CLaMP~3   & 0.163 & 0.158 & 0.197 \\
    \quad MuQ-MuLan & 0.139 & 0.112 & 0.197 \\
    \bottomrule
  \end{tabular}
\end{table}

\begin{table*}[!t]
  \centering
  \small
  \caption{DiT ablation on Song Describer (SD) and genre9k. \textbf{Base} is our free-form DiT flow model. \textbf{Null} indicates that the rewriter contains no ``Infer'' step (``Transcribe'' only). Avg.\ $\Delta$ is the mean relative gain over Base across CLAP, DSG, FAD, and MuQ. Higher is better for CLAP, DSG, MuQ, and Avg.\ $\Delta$; lower is better for FAD. \textbf{Bold}: best result per metric.}
  \label{tab:dit-ablation}
  \setlength{\tabcolsep}{2.5pt}
  \begin{tabular}{@{}ll ccccc ccccc@{}}
    \toprule
    & & \multicolumn{5}{c}{Song Describer} & \multicolumn{5}{c}{genre9k} \\
    \cmidrule(lr){3-7}\cmidrule(lr){8-12}
    DiT & Infer & CLAP $\uparrow$ & DSG $\uparrow$ & FAD $\downarrow$ & MuQ $\uparrow$ & Avg.\ $\Delta$ $\uparrow$
      & CLAP $\uparrow$ & DSG $\uparrow$ & FAD $\downarrow$ & MuQ $\uparrow$ & Avg.\ $\Delta$ $\uparrow$ \\
    \midrule
    Base & -- & 0.320 & 0.689 & 7.63 & 3.95 & 0.0\% & 0.354 & 0.744 & 7.44 & 3.93 & 0.0\% \\
    \midrule
    \multirow{2}{*}{PG} & Qwen3.5-35B & 0.338 & 0.723 & \textbf{6.59} & 3.96 & +6.2\% & 0.380 & 0.775 & \textbf{6.56} & 3.96 & +6.1\% \\
      & Null & 0.331 & 0.712 & 7.16 & 3.77 & +2.1\% & 0.372 & 0.769 & 6.89 & 3.79 & +3.1\% \\
    \midrule
    \multirow{2}{*}{PT} & Qwen3.5-35B & 0.342 & 0.735 & 7.12 & \textbf{4.02} & +5.5\% & 0.386 & 0.787 & 6.99 & \textbf{4.02} & +5.8\% \\
      & Null & 0.334 & 0.715 & 7.23 & 3.82 & +2.5\% & 0.373 & 0.767 & 7.03 & 3.84 & +2.9\% \\
    \midrule
    \multirow{2}{*}{FG} & Qwen3.5-35B & 0.344 & 0.738 & 6.68 & 3.99 & \textbf{+7.0\%} & 0.387 & 0.791 & 6.59 & 3.99 & \textbf{+7.1\%} \\
      & Null & 0.332 & 0.722 & 6.96 & 3.75 & +3.1\% & 0.376 & 0.780 & 6.71 & 3.76 & +4.2\% \\
    \midrule
    \multirow{2}{*}{FT} & Qwen3.5-35B & \textbf{0.351} & \textbf{0.756} & 7.12 & 4.00 & +6.8\% & \textbf{0.392} & \textbf{0.803} & 6.96 & 4.01 & +6.8\% \\
      & Null & 0.335 & 0.730 & 7.39 & 3.71 & +2.0\% & 0.379 & 0.781 & 7.06 & 3.76 & +3.2\% \\
    \bottomrule
  \end{tabular}
\end{table*}

\vspace{-4mm}
\section{Results}
\label{sec:results}
\subsection{Music DSG evaluation}
\label{sec:dsg-eval}

Table~\ref{tab:dsg-correlation} compares each metric's Spearman correlation with human judgments on our Internal dataset and MusicEval. Our confidence-based Music DSG outperforms all baseline metrics in all three settings. It also consistently outperforms the binary version of Music DSG, showing the value of confidence scores. The gain is especially strong for the professional panel of our Internal dataset, where DSG-conf reaches 0.414 compared with 0.385 for the next-best method.

\vspace{-2.5mm}
\subsection{DiT ablation}
\label{sec:results_dit_ablation}

Table~\ref{tab:dit-ablation} shows that full SCAPs generally benefit both training and inference. With Qwen3.5-35B completion, FT and FG consistently outperform their partial counterparts PT and PG in Avg.~$\Delta$ particularly on the two alignment metrics, CLAP and DSG.
Completion also consistently improves over leaving the same prompt-derived SCAP partial (``Null''). TSCAPs (FT, PT) yield stronger audio-text alignment, possibly because they better describe local content, while GSCAPs might include global information not presented locally. Global SCAPs (FG, PG), on the other hand, give better FAD. We hypothesize this is because the rewriter generates SCAPs for short and complete 32s music generation, thus its output is closer to the training distribution of the FG/PG DiTs. MuQ shows a mixed quality trend. We therefore use FG as the default DiT because it has the highest Avg.~$\Delta$ on both datasets.
\vspace{-2.5mm}
\subsection{Impact of rewriter configurations}
\label{sec:results_rewriter_ablation}

To study the effects from the choice of \textbf{Infer} model, we compare options on the FG DiT using Song Describer in Table~\ref{tab:rewriter-ablation}; \textbf{Transcribe} defaults to Qwen3.5-35B, except in the Claude and GPT rows, where the same model performs both stages. Every LLM-based \textbf{Infer} method improves all four metrics over Base. Among the open-source models, Qwen3.5-112B performs best overall, achieving the highest CLAP, DSG, and MuQ scores and a $+7.5\%$ average gain, while Qwen3.5-35B obtains the best FAD. GPT-5.6 Sol provides the strongest closed-source result and the highest CLAP score overall. The multimodal models also improve over Base, although less than the text-only LLMs. In contrast, \textbf{Random} inference degrades both alignment metrics and yields a $-0.2\%$ average change, showing that the gains arise from completing fields with meaningful, musically compatible values rather than merely supplying a full SCAP.

\begin{table}[h]
  \centering
  \small
  \caption{Infer model comparison on the FG DiT (Song Describer). \textbf{Bold}: best open-source model per metric; \underline{underline}: closed-source model (Claude/GPT-5.6 Sol) when it outperforms the best open-source model.}
  \label{tab:rewriter-ablation}
  \setlength{\tabcolsep}{2.5pt}
  \begin{tabular}{@{}l ccccc@{}}
    \toprule
    Infer & CLAP $\uparrow$ & DSG $\uparrow$ & FAD $\downarrow$ & MuQ $\uparrow$ & Avg.\ $\Delta$ $\uparrow$ \\
    \midrule
    \multicolumn{6}{@{}l}{\emph{Closed-source LLMs}} \\
    \quad Claude Opus 4.8 & 0.347 & 0.744 & 6.88 & 3.98 & +6.8\% \\
    \quad GPT-5.6 Sol     & \underline{0.351} & \underline{0.748} & 6.85 & 4.00 & +7.5\% \\
    \midrule
    \multicolumn{6}{@{}l}{\emph{Open-source LLMs}} \\
    \quad Qwen3.5-35B     & 0.344 & 0.738 & \textbf{6.68} & 3.99 & +7.0\% \\
    \quad Qwen3.5-112B    & \textbf{0.349} & \textbf{0.747} & 6.80 & \textbf{4.01} & \textbf{+7.5\%} \\
    \quad Qwen3.5-397B    & 0.347 & 0.746 & 6.98 & 3.99 & +6.6\% \\
    \midrule
    \multicolumn{6}{@{}l}{\emph{Multimodal LLMs}} \\
    \quad Qwen3-Omni-30B  & 0.345 & 0.738 & 6.90 & 4.00 & +6.4\% \\
    \quad FireRed Audio   & 0.342 & 0.740 & 6.97 & 3.97 & +5.8\% \\
    \midrule
    \multicolumn{6}{@{}l}{\emph{Baselines}} \\
    \quad Random          & 0.298 & 0.657 & 6.90 & 3.99 & $-0.2\%$ \\
    \bottomrule
  \end{tabular}
\end{table}

\vspace{-5.5mm}
\section{Conclusion}
\label{sec:conclusion}

We introduced SCAP-conditioned training for text-to-music generation. We investigate caption structure, attribute coverage, and temporal granularity as independent design choices and empirically verify that all SCAP-conditioned variants improve alignment and quality over free-form conditioning. While temporal and global supervision expose a tradeoff between text alignment and audio quality, training with full attribute coverage generally strengthens both axes. A multi-stage inference-time prompt-to-SCAP rewriter further improves performance by first partially filling the SCAP with information strictly adhered to user intent and subsequently populating fields with musically compatible information. Finally, the proposed Music DSG presents fine-grained attribute-level interpretations and demonstrates stronger correlation with human preference than conventional metrics. These results establish caption design, prompt rewriting, and interpretable attribute-level evaluation as complementary components for text-to-music generation.

\vfill\pagebreak

\bibliographystyle{IEEEbib}
\bibliography{refs}

@article{musiclm,
  title   = {{MusicLM}: Generating Music From Text},
  author  = {Agostinelli, A. and Denk, T. I. and Borsos, Z. and
             Engel, J. and Verzetti, M. and Caillon, A. and others},
  journal = {arXiv preprint arXiv:2301.11325},
  year    = {2023}
}

@inproceedings{musicgen,
  title     = {Simple and Controllable Music Generation},
  author    = {Copet, J. and Kreuk, F. and Gat, I. and Remez, T. and
               Kant, D. and Synnaeve, G. and others},
  booktitle = {NeurIPS},
  year      = {2023}
}

@inproceedings{sao,
  title     = {Stable Audio Open},
  author    = {Evans, Z. and Parker, J. D. and Carr, C. J. and
               Zukowski, Z. and Taylor, J. and Pons, J.},
  booktitle = {ICASSP},
  year      = {2025},
  doi       = {10.1109/ICASSP49660.2025.10888461}
}

@article{sao3,
  title         = {Stable Audio 3},
  author        = {Evans, Z. and Parker, J. D. and Rice, M. and
                   Carr, C. J. and Zukowski, Z. and Taylor, J. and others},
  journal       = {arXiv preprint arXiv:2605.17991},
  year          = {2026},
  eprint        = {2605.17991},
  archivePrefix = {arXiv},
  primaryClass  = {cs.SD}
}

@article{qwenmusic,
  title         = {{Qwen-Music} Technical Report},
  author        = {Xu, J. and Wang, K. and Yuan, R. and Lei, S. and Wang, X. and Cheng, X. and others},
  journal       = {arXiv preprint arXiv:2607.11699},
  year          = {2026},
  eprint        = {2607.11699},
  archivePrefix = {arXiv},
  primaryClass  = {cs.SD}
}

@article{qwen25omni,
  title         = {{Qwen2.5-Omni} Technical Report},
  author        = {Xu, J. and Guo, Z. and He, J. and Hu, H. and
                   He, T. and Bai, S. and others},
  journal       = {arXiv preprint arXiv:2503.20215},
  year          = {2025},
  doi           = {10.48550/arXiv.2503.20215},
  eprint        = {2503.20215},
  archivePrefix = {arXiv},
  primaryClass  = {cs.CL}
}

@inproceedings{voicebox,
  title     = {{Voicebox}: Text-Guided Multilingual Universal Speech
               Generation at Scale},
  author    = {Le, M. and Vyas, A. and Shi, B. and Karrer, B. and
               Sari, L. and Moritz, R. and others},
  booktitle = {NeurIPS},
  year      = {2023}
}

@inproceedings{musicflow,
  title     = {{MusicFlow}: Cascaded Flow Matching for Text Guided
               Music Generation},
  author    = {Prajwal, K. R. and Shi, B. and Le, M. and Vyas, A. and
               Tjandra, A. and Luthra, M. and others},
  booktitle = {ICML},
  year      = {2024}
}

@inproceedings{yue,
  title     = {{YuE}: Scaling Open Foundation Models for Long-Form
               Music Generation},
  author    = {Yuan, R. and others},
  booktitle = {ICLR},
  year      = {2026},
  url       = {https://openreview.net/forum?id=hZy6YG2Ij8}
}

@article{seed-music,
  title   = {{Seed-Music}: A Unified Framework for High Quality and
             Controlled Music Generation},
  author  = {Bai, Y. and others},
  journal = {arXiv preprint arXiv:2409.09214},
  year    = {2024},
  doi     = {10.48550/arXiv.2409.09214}
}

@article{acestep15,
  title         = {{ACE-Step 1.5}: Pushing the Boundaries of Open-Source
                   Music Generation},
  author        = {Gong, J. and Song, Y. and Zhao, W. and Wang, S. and
                   Xu, S. and Guo, J. and others},
  journal       = {arXiv preprint arXiv:2602.00744},
  year          = {2026},
  doi           = {10.48550/arXiv.2602.00744},
  eprint        = {2602.00744},
  archivePrefix = {arXiv},
  primaryClass  = {cs.SD}
}

@techreport{dalle3,
  title       = {Improving Image Generation with Better Captions},
  author      = {Betker, J. and Goh, G. and Jing, L. and Brooks, T. and
                 Wang, J. and Li, L. and others},
  institution = {OpenAI},
  year        = {2023},
  note        = {\url{https://cdn.openai.com/papers/dall-e-3.pdf}},
  url         = {https://cdn.openai.com/papers/dall-e-3.pdf}
}

@inproceedings{pixart-alpha,
  title     = {{PixArt}-$\alpha$: Fast Training of Diffusion Transformer
               for Photorealistic Text-to-Image Synthesis},
  author    = {Chen, J. and Yu, J. and Ge, C. and Yao, L. and
               Xie, E. and Wu, Y. and others},
  booktitle = {ICLR},
  year      = {2024}
}

@inproceedings{laion-clap,
  title     = {Large-Scale Contrastive Language-Audio Pretraining with
               Feature Fusion and Keyword-to-Caption Augmentation},
  author    = {Wu, Y. and Chen, K. and Zhang, T. and Hui, Y. and
               Berg-Kirkpatrick, T. and Dubnov, S.},
  booktitle = {ICASSP},
  year      = {2023},
  doi       = {10.1109/ICASSP49357.2023.10095969}
}

@article{muq-mulan,
  title   = {{MuQ}: Self-Supervised Music Representation Learning with
             Mel Residual Vector Quantization},
  author  = {Zhu, H. and Zhou, Y. and Chen, H. and Yu, J. and
             Ma, Z. and Gu, R. and others},
  journal = {arXiv preprint arXiv:2501.01108},
  year    = {2025}
}

@inproceedings{fad,
  title     = {Fr\'echet Audio Distance: A Reference-Free Metric for
               Evaluating Music Enhancement Algorithms},
  author    = {Kilgour, K. and Zuluaga, M. and Roblek, D. and Sharifi, M.},
  booktitle = {Interspeech},
  year      = {2019}
}

@article{musiceval,
  title   = {{MusicEval}: A Generative Music Corpus with Expert Ratings
             for Automatic Text-to-Music Evaluation},
  author  = {Liu, C. and Wang, H. and Zhao, J. and Zhao, S. and
             Bu, H. and Xu, X. and others},
  journal = {arXiv preprint arXiv:2501.10811},
  year    = {2025},
  url     = {https://huggingface.co/datasets/BAAI/MusicEval}
}

@article{cmi,
  title   = {{CMI-RewardBench}: Evaluating Music Reward Models with
             Compositional Multimodal Instruction},
  author  = {Ma, Y. and Xia, H. and Gao, H. and Chen, W. and
             Ye, Y. and Yang, Y. and others},
  journal = {arXiv preprint arXiv:2603.00610},
  year    = {2026}
}

@article{aqascore,
  title   = {{AQAScore}: Evaluating Semantic Alignment in Text-to-Audio
             Generation via Audio Question Answering},
  author  = {Kuan, C.-Y. and Chang, K.-W. and Lee, H.-Y.},
  journal = {arXiv preprint arXiv:2601.14728},
  year    = {2026}
}

@article{anyaudio,
  title   = {{AnyAudio-Judge}: A Dynamic Rubric-Based Benchmark and
             Evaluator for Audio Instruction Following},
  author  = {Li, H. and Tan, T. and Yang, Y. and Yang, S. and Chen, X.},
  journal = {arXiv preprint arXiv:2606.03116},
  year    = {2026}
}

@article{audioscope-tta,
  title   = {{AudioScape-TTA}: A Structured Soundscape Benchmark for
             Fine-Grained Text-to-Audio Evaluation},
  author  = {Wang, J. and Yang, Y. and Li, S. and Rong, Y. and
             Yang, S. and Yang, X. and others},
  journal = {arXiv preprint arXiv:2608.04479},
  year    = {2026}
}

@inproceedings{mustango,
  title     = {{Mustango}: Toward Controllable Text-to-Music Generation},
  author    = {Melechovsky, J. and Guo, Z. and Ghosal, D. and
               Majumder, N. and Herremans, D. and Poria, S.},
  booktitle = {NAACL-HLT},
  year      = {2024},
  doi       = {10.18653/v1/2024.naacl-long.459},
  url       = {https://aclanthology.org/2024.naacl-long.459/}
}

@inproceedings{futga,
  title     = {{FUTGA}: Towards Fine-grained Music Understanding through
               Temporally-enhanced Generative Augmentation},
  author    = {Wu, J. and Novack, Z. and Namburi, A. and Dai, J. and
               Dong, H.-W. and Xie, Z. and others},
  booktitle = {NLP4MusA},
  year      = {2024},
  url       = {https://aclanthology.org/2024.nlp4musa-1.17/}
}

@inproceedings{song-describer,
  title     = {The Song Describer Dataset: A Corpus of Audio Captions for
               Music-and-Language Evaluation},
  author    = {Manco, I. and Weck, B. and Doh, S. and Won, M. and Zhang, Y. and Bogdanov, D. and others},
  booktitle = {ML4Audio @ NeurIPS},
  year      = {2023}
}

@inproceedings{dsg,
  title     = {Davidsonian Scene Graph: Improving Reliability in
               Fine-grained Evaluation for Text-to-Image Generation},
  author    = {Cho, J. and Hu, Y. and Garg, R. and Anderson, P. and
               Krishna, R. and Baldridge, J. and others},
  booktitle = {ICLR},
  year      = {2024}
}

@inproceedings{tifa,
  title     = {{TIFA}: Accurate and Interpretable Text-to-Image
               Faithfulness Evaluation with Question Answering},
  author    = {Hu, Y. and Liu, B. and Kasai, J. and Wang, Y. and
               Ostendorf, M. and Krishna, R. and others},
  booktitle = {ICCV},
  year      = {2023}
}

@inproceedings{music-metadata-llm,
  title     = {Rethinking Music Captioning with Music Metadata {LLMs}},
  author    = {Bukey, I. and Wang, Z. and Donahue, C. and Bryan, N. J.},
  booktitle = {ICASSP},
  year      = {2026}
}

@inproceedings{rectified-flow,
  title     = {Flow Straight and Fast: Learning to Generate and Transfer
               Data with Rectified Flow},
  author    = {Liu, X. and Gong, C. and Liu, Q.},
  booktitle = {ICLR},
  year      = {2023},
  url       = {https://openreview.net/forum?id=XVjTT1nw5z}
}

@inproceedings{scale-rectified-flow,
  title     = {Scaling Rectified Flow Transformers for High-Resolution
               Image Synthesis},
  author    = {Esser, P. and Kulal, S. and Blattmann, A. and Entezari, R. and
               M\"{u}ller, J. and Saini, H. and others},
  booktitle = {ICML},
  year      = {2024},
  url       = {https://proceedings.mlr.press/v235/esser24a.html}
}

@inproceedings{skey,
  title     = {{S-KEY}: Self-supervised Learning of Major and Minor Keys
               from Audio},
  author    = {Kong, Y. and Meseguer-Brocal, G. and Lostanlen, V. and
               Lagrange, M. and Hennequin, R.},
  booktitle = {ICASSP},
  year      = {2025}
}

@inproceedings{davies2019tcn,
  title     = {Temporal Convolutional Networks for Musical Audio Beat
               Tracking},
  author    = {Davies, M. E. P. and B{\"o}ck, S.},
  booktitle = {EUSIPCO},
  year      = {2019}
}

@inproceedings{bock2016joint,
  title     = {Joint Beat and Downbeat Tracking with Recurrent Neural
               Networks},
  author    = {B{\"o}ck, S. and Krebs, F. and Widmer, G.},
  booktitle = {ISMIR},
  year      = {2016}
}

@article{ghosh2025music,
  title         = {Music Flamingo: Scaling Music Understanding in Audio
                   Language Models},
  author        = {Ghosh, S. and Goel, A. and Koroshinadze, L. and
                   Lee, S.-G. and Kong, Z. and Santos, J. F. and others},
  journal       = {arXiv preprint arXiv:2511.10289},
  year          = {2025},
  doi           = {10.48550/arXiv.2511.10289},
  eprint        = {2511.10289},
  archivePrefix = {arXiv},
  primaryClass  = {eess.AS},
  url           = {https://arxiv.org/abs/2511.10289}
}

@INPROCEEDINGS{stae,
  author={Casebeer, J. and Zhu, G. and Wang, Z. and Bryan, N. J.},
  booktitle={ICASSP}, 
  title={A Generative-First Neural Audio Autoencoder}, 
  year={2026},
  doi={10.1109/ICASSP55912.2026.11463285}
}

@inproceedings{doh2026llmjudge,
  author    = {Doh, S. and Sguerra, B. and Oramas, S. and Epure, E. V. and Nam, J.},
  title     = {{LLM-as-a-Judge} for Evaluating System Responses in Conversational Music Recommendation},
  booktitle = {RecSys},
  year      = {2026},
}

\end{document}